\documentclass[preprint,aps,showpacs,showkeys,nofootinbib]{revtex4}
\usepackage{graphicx}
\usepackage{dcolumn}
\usepackage{bm}
\begin{document}
\title{Muon conversion to electron in nuclei within the BLMSSM}

\author{Tao Guo$^{1,2}$\footnote{sjzueguotao@163.com},
Shu-Min Zhao$^{3}$\footnote{zhaosm@hbu.edu.com},
Xing-Xing Dong$^{3}$\footnote{dxx\_0304@163.com}, Chun-Gui Duan$^{1}$\footnote{duancg@mail.hebtu.edu.cn},
Tai-Fu Feng$^{3}$\footnote{fengtf@hbu.edu.cn}
}

\affiliation{$^1$ Department of Physics and Hebei Advanced Thin Films Laboratory, Hebei Normal University, Shijiazhuang 050024, China\\
$^2$ College of Mathematics and Physics, Hebei GEO University, Shijiazhuang 050031, China\\
$^3$ Department of Physics and Technology, Hebei University, Baoding 071002, China}
\date{\today}
\begin{abstract}
In a supersymmetric extension
of the standard model with local gauged baryon
and lepton numbers (BLMSSM), there are new sources for lepton flavor violation, because the right-handed neutrinos, new gauginos and Higgs are
introduced. We investigate muon conversion to electron in nuclei within the BLMSSM in detail.
The numerical results indicate that the $\mu \rightarrow e $ conversion
rates in nuclei within the BLMSSM can reach the experimental upper bound, which may be
detected in the future experiments.
\end{abstract}

\maketitle
\section{Introduction}

The observations of neutrino oscillations \cite{neutrino1,neutrino2,neutrino3} imply that neutrinos have
tiny masses and are mixed \cite{neutrino4,neutrino5,neutrino6,neutrino7}, which have demonstrated that lepton flavor in neutrino
sector is not conserved. Nevertheless, in the Standard Model (SM) with three tiny
massive neutrinos, the expected rates for the charged lepton flavor violating (LFV)
processes are very tiny\cite{LFV1,LFV2,LFV3,LFV4}. Thus, Lepton-flavor violation is a window of new physics beyond the SM.
 Among the various
candidates for new physics that produce potentially observable effects in LFV processes,
one of the most appealing model is supersymmetric (SUSY) extension of the SM. Here, we can use the neutrino oscillation experimental data to restrain the input parameters in the new models. A neutral Higgs with mass $m_{h^0}= 125.1$ GeV reported
by ATLAS \cite{LHC1} and CMS \cite{LHC2,LHC3} gives a strict constraint on relevant parameter
space of the model.

The present sensitivities of the $\mu -e$ conversion rates
in different nuclei \cite{experiment1,experiment2,experiment3} are collected here,
\begin{eqnarray}
&&{\rm CR}(\mu\rightarrow e:{\rm Au})<7  \times10^{-13},\nonumber\\
&&{\rm CR}(\mu\rightarrow e:{\rm Ti})<4.3\times10^{-12},\nonumber\\
&&{\rm CR}(\mu\rightarrow e:{\rm Pb})<4.6\times10^{-11}.
\end{eqnarray}
These processes have close relation with $l_j\rightarrow l_i\gamma$. In the work\cite{ZHB}, the $\mu\rightarrow e$ conversion was studied in $\mu\nu$SSM.
For the models beyond SM, one can violate R parity with the non-conservation of
baryon number ($B$) or lepton number ($L$)\cite{BLMSSM1,BLMSSM2,Rp1,Rp2}.
A minimal supersymmetric extension of
the SM with local gauged $B$ and $L$(BLMSSM) was first proposed by the author\cite{BLMSSM3,BLMSSM4}.
The local gauged $B$ is used to explain the matter-antimatter asymmetry in the universe.
Right-handed neutrinos in BLMSSM lead to three tiny neutrino masses
through the See-Saw mechanism and can account for the neutrino oscillation experiments.
So lepton number ($L$) is expected to be broken spontaneously around ${\rm TeV}$ scale.

In BLMSSM, the lightest CP-even Higgs mass and the decays $h^0\rightarrow\gamma\gamma$, $h^0\rightarrow ZZ (WW)$ were studied
in the work\cite{weBLMSSM}. The
neutron and lepton electric dipole moments(EDMs) were researched in the CP-violating BLMSSM\cite{smneutron,BLLEDM}.
In BLMSSM, there are also other works \cite{weBLMSSM1,weBLMSSM2,weBLMSSM3}.
In this work, we analyze the processes on muon conversion
to electron in nuclei within the BLMSSM. Compared with MSSM, there
are new sources to enlarge the processes via loop contributions. The new  scores are produced from:
1. the coupling of new neutralino(lepton neutralino)-slepton-lepton;
2. the right-handed neutrinos mixing with left-handed neutrinos;
3. the sneutrino sector is extended, whose mass squared matrix is $6\times 6$.
In some parameter space of BLMSSM, large corrections to the processes are obtained, and they can easily exceed their experiment upper bounds.
Therefore, to enhance the processes on muon conversion
to electron in nuclei is possible, and they may be measured in the near future.

After this introduction, we briefly summarize the main ingredients
of the BLMSSM, and show the needed mass matrices and couplings in section II.
In section III, the  processes $\mu\rightarrow \emph{e}+\rm{q}\bar q$
are studied in the BLMSSM.
The input parameters and numerical analysis are shown in section IV, and our conclusion is given in section V.
Some functions are collected in the Appendix.

\section{BLMSSM}
BLMSSM is the supersymmetric extension of the SM with local gauged $B$ and $L$, whose local gauge group is $SU(3)_{C}\otimes SU(2)_{L}\otimes U(1)_{Y}\otimes U(1)_{B}\otimes U(1)_{L}$\cite{BLMSSM1,BLMSSM2}.
The exotic leptons $\hat{L}_{4}$, $\hat{E}_{4}^c$, $\hat{N}_{4}^c$, $\hat{L}_{5}^c$, $\hat{E}_{5}$ and $\hat{N}_{5}$ are introduced  to cancel $L$ anomaly. As well as,
the exotic quarks $\hat{Q}_{4}$, $\hat{U}_{4}^c$, $\hat{D}_{4}^c$, $\hat{Q}_{5}^c$, $\hat{U}_{5}$ and $\hat{D}_{5}$ are introduced  to cancel $B$ anomaly.
To break lepton number and baryon number spontaneously, the Higgs superfields $\hat{\Phi}_{L}$, $\hat{\varphi}_{L}$  and
$\hat{\Phi}_{B}$, $\hat{\varphi}_{B}$ are introduced, respectively.
The exotic quarks obtain masses from nonzero vacuum expectation values(VEVs) of $\hat{\Phi}_B$ and $\hat{\varphi}_B$. While, exotic leptons get masses from VEVs of $H_u$ and $H_d$.
$H_u$ and $\hat{\varphi}_L$ give masses to light neutrinos through See-Saw mechanism.
In the BLMSSM, the superfields $\hat{X}$ and
$\hat{X}^\prime$ are introduced to make the heavy exotic quarks unstable. The above mentioned exotic lepton, quark and Higgs superfields are shown in table 1.

\begin{table}[h]
\caption{ \label{1}  Superfields in the BLMSSM.}
\begin{tabular*}{87.5mm}{|c|c|c|c|c|c|}
\hline
Superfields &$SU(3)_C$ &$SU(2)_L$ &$U(1)_Y$ &$U(1)_B$ &$U(1)_L$ \\
\hline
$\hat{L}_4$&1 &2 &-1/2 &0 &$L_4$ \\
\hline
$\hat{E}_4^c$&1 &1 &1 &0 &-$L_4$ \\
\hline
$\hat{N}_4^c$&1 &1 &0 &0 &-$L_4$ \\
\hline
$\hat{L}_5^c$&1 &2 &1/2 &0 &-3-$L_4$ \\
\hline
$\hat{E}_5$&1 &1 &-1 &0 &3+$L_4$ \\
\hline
$\hat{N}_5$&1 &1 &0 &0 &3+$L_4$ \\
\hline
$\hat{Q}_4$&3 &2 &1/6 &$B_4$ &0 \\
\hline
$\hat{U}_4^c$&$\bar{3}$ &1 &-2/3 &-$B_4$ &0 \\
\hline
$\hat{D}_4^c$&$\bar{3}$ &1 &1/3 &-$B_4$ &0 \\
\hline
$\hat{Q}_5^c$&$\bar{3}$ &2 &-1/6 &-1-$B_4$ &0 \\
\hline
$\hat{U}_5$&3 &1 &2/3 &1+$B_4$ &0 \\
\hline
$\hat{D}_5$&3 &1 &-1/3 &1+$B_4$ &0 \\
\hline
$\hat{\Phi}_L$&1 &1 &0 &0 &-2 \\
\hline
$\hat{\varphi}_L$&1 &1 &0 &0 &2 \\
\hline
$\hat{\Phi}_B$&1 &1 &0 &1 &0 \\
\hline
$\hat{\varphi}_B$&1 &1 &0 &-1 &0 \\
\hline
$\hat{X}$&1 &1 &0 &2/3+$B_4$ &0 \\
\hline
$\hat{X}'$&1 &1 &0 &-2/3-$B_4$ &0 \\
\hline
\end{tabular*}%
\end{table}

The superpotential of BLMSSM is\cite{weBLMSSM}
\begin{eqnarray}
&&{\cal W}_{{BLMSSM}}={\cal W}_{{MSSM}}+{\cal W}_{B}+{\cal W}_{L}+{\cal W}_{X}\;,
\label{superpotential1}
\nonumber\\&&{\cal W}_{B}=\lambda_{Q}\hat{Q}_{4}\hat{Q}_{5}^c\hat{\Phi}_{B}+\lambda_{U}\hat{U}_{4}^c\hat{U}_{5}
\hat{\varphi}_{B}+\lambda_{D}\hat{D}_{4}^c\hat{D}_{5}\hat{\varphi}_{B}+\mu_{B}\hat{\Phi}_{B}\hat{\varphi}_{B}
\nonumber\\
&&\hspace{1.2cm}
+Y_{{u_4}}\hat{Q}_{4}\hat{H}_{u}\hat{U}_{4}^c+Y_{{d_4}}\hat{Q}_{4}\hat{H}_{d}\hat{D}_{4}^c
+Y_{{u_5}}\hat{Q}_{5}^c\hat{H}_{d}\hat{U}_{5}+Y_{{d_5}}\hat{Q}_{5}^c\hat{H}_{u}\hat{D}_{5}\;,
\nonumber\\
&&{\cal W}_{L}=Y_{{e_4}}\hat{L}_{4}\hat{H}_{d}\hat{E}_{4}^c+Y_{{\nu_4}}\hat{L}_{4}\hat{H}_{u}\hat{N}_{4}^c
+Y_{{e_5}}\hat{L}_{5}^c\hat{H}_{u}\hat{E}_{5}+Y_{{\nu_5}}\hat{L}_{5}^c\hat{H}_{d}\hat{N}_{5}
\nonumber\\
&&\hspace{1.2cm}
+Y_{\nu}\hat{L}\hat{H}_{u}\hat{N}^c+\lambda_{{N^c}}\hat{N}^c\hat{N}^c\hat{\varphi}_{L}
+\mu_{L}\hat{\Phi}_{L}\hat{\varphi}_{L}\;,
\nonumber\\
&&{\cal W}_{X}=\lambda_1\hat{Q}\hat{Q}_{5}^c\hat{X}+\lambda_2\hat{U}^c\hat{U}_{5}\hat{X}^\prime
+\lambda_3\hat{D}^c\hat{D}_{5}\hat{X}^\prime+\mu_{X}\hat{X}\hat{X}^\prime\;.
\label{superpotential-BL}
\end{eqnarray}
where ${\cal W}_{{MSSM}}$ is the superpotential of the MSSM.
To save space in the text, the soft breaking terms $\mathcal{L}_{{soft}}$\cite{BLMSSM1, weBLMSSM} of the BLMSSM is not shown here.

In this model, we introduce the superfields $\hat{N}^c$ to produce tiny masses of three light neutrinos. The mass matrix of neutrinos
in the basis $(\psi_{\nu_L^I}, \psi_{N_R^{cI}})$ is expressed as
\begin{eqnarray}
&&Z_{N_{\nu}}^\top\left(\begin{array}{cc}
  0&\frac{v_u}{\sqrt{2}}(Y_{\nu})^{IJ} \\
   \frac{v_u}{\sqrt{2}}(Y^{T}_{\nu})^{IJ}  & \frac{\bar{v}_L}{\sqrt{2}}(\lambda_{N^c})^{IJ}
    \end{array}\right)Z_{N_{\nu}}=diag(m_{\nu^{\alpha}}), \alpha=1\cdot\cdot\cdot6,I,J=1,2,3,
\nonumber\\&& \psi_{{\nu_L^I}}=Z_{{N_{\nu}}}^{I\alpha}k_{N_\alpha}^0,\;\;\;\;
\psi_{N_R^{cI}}=Z_{{N_{\nu}}}^{(I+3)\alpha}k_{N_\alpha}^0,\;\;\;\;
\chi_{N_\alpha}^0=\left(\begin{array}{c}
   k_{N_\alpha}^0\\  \bar{k}_{N_\alpha}^0
    \end{array}\right).
\end{eqnarray}
Here, $\chi_{N_\alpha}^0$ represent the mass eigenstates of neutrino fields mixed by left-handed and right-handed neutrinos.

The new gaugino $\lambda_L$ mixes with the superpartners of the $SU(2)_L$ singlets $\Phi_L$ and $\varphi_L$, then they produce three lepton neutralinos
\begin{equation}
\mathcal{L}_{\chi_L^0}=\frac{1}{2}(i\lambda_L,\psi_{\Phi_L},\psi_{\varphi_L})\left(     \begin{array}{ccc}
  2M_L &2v_Lg_L &-2\bar{v}_Lg_L\\
   2v_Lg_L & 0 &-\mu_L\\-2\bar{v}_Lg_L&-\mu_L &0
    \end{array}\right)  \left( \begin{array}{c}
 i\lambda_L \\ \psi_{\Phi_L}\\\psi_{\varphi_L}
    \end{array}\right)+h.c.\label{LN}
   \end{equation}
    One can use $Z_L$ to  diagonalize the mass matrix in Eq.(\ref{LN}) and obtain three lepton neutralino masses in the end.

From Eqs.(\ref{superpotential-BL}) and the soft breaking terms $\mathcal{L}_{{soft}}$\cite{BLMSSM1, weBLMSSM} of the BLMSSM, the mass squared matrix of slepton gets corrections and reads as
\begin{eqnarray}
&&\left(\begin{array}{cc}
  (\mathcal{M}^2_L)_{LL}&(\mathcal{M}^2_L)_{LR} \\
   (\mathcal{M}^2_L)_{LR}^{\dag} & (\mathcal{M}^2_L)_{RR}
    \end{array}\right).
\end{eqnarray}

$(\mathcal{M}^2_L)_{LL},~(\mathcal{M}^2_L)_{LR}$ and $(\mathcal{M}^2_L)_{RR}$ are shown here
\begin{eqnarray}
 &&(\mathcal{M}^2_L)_{LL}=\frac{(g_1^2-g_2^2)(v_d^2-v_u^2)}{8}\delta_{IJ} +g_L^2(\bar{v}_L^2-v_L^2)\delta_{IJ}
 +m_{l^I}^2\delta_{IJ}+(m^2_{\tilde{L}})_{IJ},\nonumber\\&&
 (\mathcal{M}^2_L)_{LR}=\frac{\mu^*v_u}{\sqrt{2}}(Y_l)_{IJ}-\frac{v_u}{\sqrt{2}}(A'_l)_{IJ}+\frac{v_d}{\sqrt{2}}(A_l)_{IJ},
 \nonumber\\&& (\mathcal{M}^2_L)_{RR}=\frac{g_1^2(v_u^2-v_d^2)}{4}\delta_{IJ}-g_L^2(\bar{v}_L^2-v_L^2)\delta_{IJ}
 +m_{l^I}^2\delta_{IJ}+(m^2_{\tilde{R}})_{IJ}.
\end{eqnarray}
The  unitary matrix $Z_{\tilde{L}}$ is used to rotate slepton mass squared matrix to mass eigenstates.

Because of the introduction of right handed neutrinos, in BLMSSM the mass squared matrix of sneutrino is $6\times 6$.
In the base $\tilde{n}^{T}=(\tilde{\nu},\tilde{N}^{c})$,
the concrete forms for the sneutrino mass squared matrix ${\cal M}_{\tilde{n}}$ are shown here
    \begin{eqnarray}
  && {\cal M}^2_{\tilde{n}}(\tilde{\nu}_{I}^*\tilde{\nu}_{J})=\frac{g_1^2+g_2^2}{8}(v_d^2-v_u^2)\delta_{IJ}+g_L^2(\overline{v}^2_L-v^2_L)\delta_{IJ}
   +\frac{v_u^2}{2}(Y^\dag_{\nu}Y_\nu)_{IJ}+(m^2_{\tilde{L}})_{IJ},\nonumber\\&&
   {\cal M}^2_{\tilde{n}}(\tilde{N}_I^{c*}\tilde{N}_J^c)=-g_L^2(\overline{v}^2_L-v^2_L)\delta_{IJ}
   +\frac{v_u^2}{2}(Y^\dag_{\nu}Y_\nu)_{IJ}+2\overline{v}^2_L(\lambda_{N^c}^\dag\lambda_{N^c})_{IJ}\nonumber\\&&
   \hspace{1.8cm}+(m^2_{\tilde{N}^c})_{IJ}+\mu_L\frac{v_L}{\sqrt{2}}(\lambda_{N^c})_{IJ}
   -\frac{\overline{v}_L}{\sqrt{2}}(A_{N^c})_{IJ}(\lambda_{N^c})_{IJ},\nonumber\\&&
   {\cal M}^2_{\tilde{n}}(\tilde{\nu}_I\tilde{N}_J^c)=\mu^*\frac{v_d}{\sqrt{2}}(Y_{\nu})_{IJ}-v_u\overline{v}_L(Y_{\nu}^\dag\lambda_{N^c})_{IJ}
   +\frac{v_u}{\sqrt{2}}(A_{N})_{IJ}(Y_\nu)_{IJ}.
   \end{eqnarray}

The superfields $\tilde{N}^c$ in BLMSSM lead to the corrections
for some couplings existed in MSSM.
We give out the corrected couplings such as: W-lepton-neutrino and Z-neutrino-neutrino
\begin{eqnarray}
&&\mathcal{L}_{WL\nu}=-\frac{e}{\sqrt{2}s_W}W_{\mu}^+\sum_{I=1}^3\sum_{\alpha=1}^6Z_{N_{\nu}}^{I\alpha*}\bar{\chi}_{N_{\alpha}}^0\gamma^{\mu}P_Le^I,
\nonumber\\&&\mathcal{L}_{Z\nu\nu}=-\frac{e}{2s_Wc_W}Z_{\mu}\sum_{I=1}^3\sum_{\alpha,\beta=1}^6Z_{N_{\nu}}^{I\alpha*}Z_{N_{\nu}}^{I\beta}\bar{\chi}_{N_{\alpha}}^0\gamma^{\mu}P_L\chi_{N_{\beta}}^0,
\end{eqnarray}
where $P_L=\frac{1-\gamma_5}{2}$ and $P_R=\frac{1+\gamma_5}{2}$. We use the abbreviation $s_W=\sin\theta_W$, $c_W=\cos\theta_W$, and $\theta_W$ is the Weinberg angle.

Some other adapted couplings are collected here: chargino-lepton-sneutrino, Z-sneutrino-sneutrino and charged Higgs-lepton-neutrino
\begin{eqnarray}
&&\mathcal{L}_{\chi^{\pm}L\tilde{\nu}}=-\sum_{I=1}^3\sum_{i=1}^6\sum_{j=1}^2\bar{\chi}^-_j
\Big(Y_l^{I} Z_-^{2j*}Z_{\nu}^{Ii*}P_R+
[\frac{e}{s_W}Z_+^{1j}Z_{\nu}^{Ii*}
\nonumber\\&&\hspace{1.8cm}+Y_\nu^{Ii}Z_+^{2j}Z_{\nu}^{(I+3)i*}]P_L
\Big)e^I\tilde{\nu}^{i*}+h.c.
\\
&&\mathcal{L}_{Z\tilde{\nu}\tilde{\nu}}=-\frac{e}{2s_Wc_W}Z_{\mu}\sum_{I=1}^3\sum_{i,j=1}^6Z_{\nu}^{Ii*}Z_{\nu}^{Ij}\tilde{\nu}^{i*}i(\overrightarrow{\partial}^{\mu}
-\overleftarrow{\partial}^{\mu})\tilde{\nu}^j.
\\
&&\mathcal{L}_{H^{\pm}L\nu}=\sum_{I}^3\sum_{\alpha=1}^6G^{\pm}
\bar{e}^I\Big(Y_l^{I}\cos\beta Z_{N_{\nu}}^{I\alpha}P_L-Y_{\nu}^{I\alpha*}\sin\beta Z_{N_{\nu}}^{(I+3)\alpha}P_R
\Big)\chi_{N_{\alpha}}^0\nonumber\\&&-\sum_{I}^3\sum_{\alpha=1}^6
H^{\pm}\bar{e}^I\Big(Y_l^{I}\sin\beta Z_{N_{\nu}}^{I\alpha}P_L+Y_{\nu}^{I\alpha*}\cos\beta Z_{N_{\nu}}^{(I+3)\alpha}P_R
\Big)\chi_{N_{\alpha}}^0+h.c.\end{eqnarray}

In BLMSSM, there are new couplings that are deduced from
the interactions of gauge and matter multiplets
$ig\sqrt{2}T^a_{ij}(\lambda^a\psi_jA_i^*-\bar{\lambda}^a\bar{\psi}_iA_j)$.
After calculation, the lepton-slepton-lepton neutralino couplings are obtained
\begin{eqnarray}
&&\mathcal{L}_{l\chi_L^0\tilde{L}}=\sum_{i=1}^6\sum_{j=1}^3\sqrt{2}g_L\bar{\chi}_{L_j}^0\Big(Z_{N_L}^{1j}Z_{L}^{Ii}P_L
-Z_{N_L}^{1j*}Z_{L}^{(I+3)i}P_R\Big)l^I\tilde{L}_i^++h.c.
\end{eqnarray}

\section{$\mu\rightarrow \emph{e}+\rm{q}\bar q$ in the BLMSSM}
In this section, the LFV processes $\mu\rightarrow \emph{e}+\rm{q}\bar q$
are studied in the BLMSSM. Both penguin-type diagrams and box-type diagrams have contributions to the effective Lagrangian. For convenience, the penguin and box diagrams are analyzed in the generic form, which can simplify
the work.
\begin{figure}
\setlength{\unitlength}{1mm}
\centering
\includegraphics[width=5.0in]{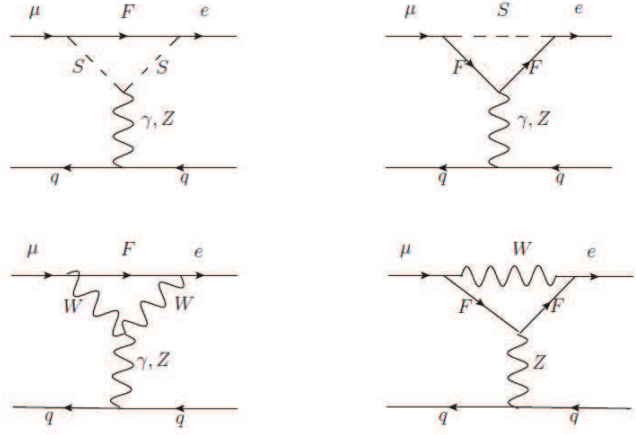}
\vspace{-10cm}
\caption[]{The penguin diagrams for $\mu\rightarrow \emph{e}+\rm{q}\bar q$, with $F$ representing Dirac(Majorana) particles.}\label{FMFLV1}
\end{figure}
\subsection{the penguin diagrams}

When the external leptons are all on shell, we can generally obtain the $\gamma$-penguin contributions in the following form
\begin{eqnarray}
&&T_{\gamma {\rm{ - p}}}  =  - \bar e\left( {p_1 } \right)\left[ {q^2 \gamma _\mu  \left( {C_1^L P_L  + C_1^R P_R } \right) + m_\mu  i\sigma _{\mu \nu } q^\nu  \left( {C_2^L P_L  + C_2^R P_R } \right)} \right]\mu \left( p \right)
\nonumber\\
&&\qquad
\times \frac{{e^2 }}{{q^2 }}\bar q\left( {p_2 } \right)\gamma ^\mu  q\left( {p_3 } \right)
\label{amplitude-gamma}
\end{eqnarray}
The relevant Feynman diagrams are shown in Fig.1. The final Wilson coefficients
$C_1^L,C_1^R,C_2^L$ and $C_2^R$ are obtained from the sum of these diagrams' amplitudes.

The contributions from the virtual neutral fermion diagram in the top-left of Fig.\ref{FMFLV1} are denoted by $C_\alpha^{L,R}(n),\alpha=1,2$.
We give out the deduced results in the following form,
\begin{eqnarray}
&&C_1^{L}(n) = \sum_{F=\chi^0,\chi_L^0,\nu}\sum_{S=\tilde{L},\tilde{L},H^{\pm}}\frac{1}{6{m_W^2}}H_R^{SF\bar{e}}H_L^{S^*\mu\bar{F}}{I_1}
({x_F},{x_S})\;,\nonumber\\
&&C_2^{L}(n) =\sum_{F=\chi^0,\chi_L^0,\nu}\sum_{S=\tilde{L},\tilde{L},H^{\pm}} \frac{{{m_F}}}{{{m_{{\mu}}}}{m_W^2}}H_L^{SF\bar{e}}H_L^{S^*\mu\bar{F}}
\Big[ {I_2}({x_F},{x_S}) - {I_3}({x_F},{x_S}) \Big]
\;, \nonumber\\
&&C_\alpha^{R}(n) = \left. {C_\alpha^{L}}(n) \right|{ _{L \leftrightarrow R}},~~~\alpha=1,2.\label{oneloopn}
\end{eqnarray}
with $x= {m^2}/{m_W^2}$ and $m$ representing the mass for the corresponding particle. $H_{L,R}^{SF\bar{e}}$ and $H_{L,R}^{S^*\mu\bar{F}}$
are the corresponding couplings of the left(right)-hand parts in the Lagrangian.
The one-loop functions $I_{i}(x_1,x_2),i=1\dots4$ are collected in Appendix.

The diagram in top-right of Fig.\ref{FMFLV1} represents the virtual charged Fermion diagram and its contribution is
\begin{eqnarray}
&&C_1^{L}(c) = \sum_{F=\chi^{\pm}}\sum_{S=\tilde{\nu}} \frac{1}{6{m_W^2}}
H_R^{SF\bar{e}}H_L^{S^*\mu \bar{F}}
\Big[ {I_3}({x_F},{x_S})- 2 {I_4}({x_F},{x_S})- {I_1}({x_F},{x_S}) \Big] \;, \nonumber\\
&&C_2^{L}(c) = \sum_{F=\chi^{\pm}}\sum_{S=\tilde{\nu}} \frac{{{m_F}}}{{{m_{{\mu}}}}{m_W^2}}
H_L^{SF\bar{e}}H_L^{S^*\mu \bar{F}}\Big[ {I_3}({x_F},{x_S}) - {I_4}({x_F},{x_S}) - {I_1}({x_F},{x_S}) \Big] \;, \nonumber\\
&&C_\alpha^{R}(c) = \left. {C_\alpha^{L}}(c) \right|{ _{L \leftrightarrow R}}, ~~~\alpha=1,2.\label{oneloopc}
\end{eqnarray}

On account of the mixing of three light neutrinos and three heavy neutrinos, the virtual $W$ diagrams in the bottom of Fig.\ref{FMFLV1} give
corrections to the charged LFV process $\mu\rightarrow \emph{e}+\rm{q}\bar q$. We show the coefficients
$C_\alpha^{L,R}(W)(\alpha=1,2)$
\begin{eqnarray}
&&C^L_1(W)=\sum_{F=\nu}\frac{-1}{2m_W^2}H^{WF\bar{e}}_LH^{W^*\mu\bar{F}}_L\Big[I_2(x_{F},x_W)+I_1(x_{F},x_W)\Big],
\nonumber\\
&&C^L_2(W)=\sum_{F=\nu}\frac{1}{m_W^2}H^{WF\bar{e}}_LH^{W^*\mu\bar{F}}_L(1+\frac{m_{e}}{m_{\mu}})
\Big[2I_2(x_{F},x_W)-\frac{1}{3}I_1(x_{F},x_W)\Big],
\nonumber\\&&C_\alpha^{R}(W)=0,~~~\alpha=1,2.\label{oneloopW}
\end{eqnarray}
The sum of the total coefficients in Eqs.(\ref{oneloopn})(\ref{oneloopc})(\ref{oneloopW}) are
\begin{eqnarray}
C_\alpha^{L,R}=C_\alpha^{L,R}(n)+C_\alpha^{L,R}(c)+C_\alpha^{L,R}(W),~~~\alpha=1,2.
\end{eqnarray}

The contributions from $Z$-penguin diagrams are depicted by the Fig.1, similar as $\gamma$-penguin diagrams,
\begin{eqnarray}
&&T_{Z- {\rm{p}}} = \frac{{{e^2}}}{{m_Z^2}}{\bar e}({p_1}){\gamma _\mu }
({N_L}{P_L} + {N_R}{P_R}){\mu}(p)  {\bar q}({p_2}){\gamma ^\mu }\Big(H_L^{Ze{\bar{e}}}{P_L} \qquad \nonumber\\
&&\qquad\quad\;\; + \:H_R^{Ze{\bar{e}}}{P_R}\Big){q}({p_3}) \:,
\nonumber\\&&{N_{L,R}} = N_{L,R}(S) +N_{L,R}(W)\:.
\end{eqnarray}

The concrete forms of the effective couplings $N_{L}(S),~N_{R}(S),~N_L(W)$ and $N_R(W)$ read as
\begin{eqnarray}
&&N_L(S) = \frac{1}{2{e^2}}\sum\limits_{F=\chi^0,\chi^{\pm},\nu}
 \sum\limits_{S=\tilde{L},\tilde{\nu},H^{\pm}}\Big[\frac{2{m_{F_1}}{m_{F_2}}}
{{m_W^2}}H_R^{SF_2{{\bar e }}}H_L^{ZF_1{{\bar F }_2 }}
H_L^{S^*{\mu}{{\bar F }_1 }}{G_1}({x_S},{x_{F_2}},{x_{F_1}}) \nonumber\\&&\hspace{1.2cm}+H_R^{S_2 F{{\bar e }}}
H_R^{ZS_1 S_2^ {*} }H_L^{S_1 ^{*} {\mu}\bar F}{G_2}
({x_F},{x_{S_1 }},{x_{S_2}})-  H_R^{SF_2{{\bar e }}}
H_R^{ZF_1{{\bar F }_2 }}H_L^{S^*{\mu}{{\bar F }_1 }}{G_2}({x_S},{x_{F_2}},{x_{F_1}})
\Big]\nonumber\\&&\hspace{1.2cm}+\sum\limits_{F=\chi^0_L}
 \sum\limits_{S=\tilde{L}}\Big[H_R^{S_2 F{{\bar e }}}
H_R^{ZS_1 S_2^ {*} }H_L^{S_1 ^{*} {\mu}\bar F}{G_2}
({x_F},{x_{S_1 }},{x_{S_2}})\Big],\nonumber\\
&&N_R(S) = \left. {N_L}(S) \right|{ _{L \leftrightarrow R}},
\nonumber\\&&N_L(W)=\frac{c_W}{es_W}\sum_{F=\nu}H^{WF\bar{e}}_LH^{W^*\mu\bar{F}}_L
\Big[G_3(x_F,x_W)+2(x_i+x_j)[I_1(x_F,x_W)-I_2(x_F,x_W)]\Big]\nonumber\\&&\hspace{1.8cm}
+\frac{1}{e^2}\sum_{F_1,F_2=\nu}H^{WF_2\bar{e}}_LH^{W^*\mu\bar{F}_1}_LH^{Z^*F_1\bar{F}_2}_L
\Big(-\frac{3}{32\pi^2}-G_2(x_W,x_{F_1},x_{F_2})\nonumber\\&&\hspace{1.8cm}+x_j[\frac{1}{3}G_4(x_W,x_{F_1},x_{F_2})
+G_5(x_W,x_{F_1},x_{F_2})]
\Big),\nonumber\\&&
N_R(W)=\frac{c_W}{es_W}\sum_{F=\nu}H^{WF\bar{e}}_LH^{W^*\mu\bar{F}}_L\Big[2\sqrt{x_ix_j}[I_1(x_F,x_W)-I_2(x_F,x_W)]\Big]
\nonumber\\&&\hspace{1.8cm}+\frac{1}{e^2}\sum_{F_1,F_2=\nu}H^{WF_2\bar{e}}_LH^{W^*\mu\bar{F}_1}_LH^{Z^*F_1\bar{F}_2}_L\sqrt{x_ix_j}
\Big(2G_1(x_W,x_{F_1},x_{F_2})\nonumber\\&&\hspace{1.8cm}
-\frac{1}{3}G_4(x_W,x_{F_1},x_{F_2})
-2G_5(x_W,x_{F_1},x_{F_2})  \Big).
\end{eqnarray}
The concrete expressions for the functions $G_i$  $(i=1,...,7)$ are collected are in appendix.

\subsection{The box-type diagrams}
\begin{figure}
\setlength{\unitlength}{1mm}
\centering
\includegraphics[width=5.8in]{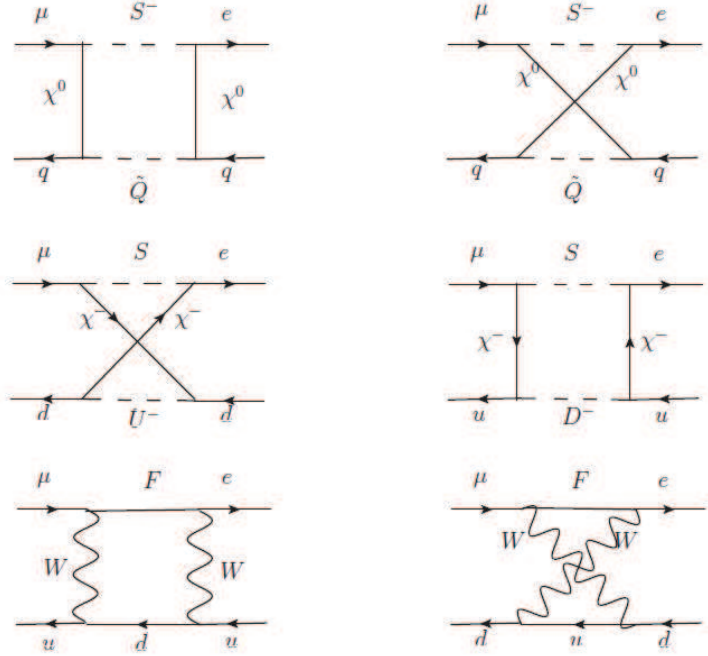}
\vspace{-10cm}
\caption[]{The box diagrams for $\mu\rightarrow \emph{e}+\rm{q}\bar q$, with $F$ representing Dirac(Majorana) particles.}\label{FMFLV2}
\end{figure}

The box-type diagrams drawn in Fig.\ref{FMFLV2}  can be written as

\begin{eqnarray}
&& T_{{\rm{box}}}  = e^2 \sum_{q=u,d}\bar q\gamma _\alpha  q\bar e\gamma ^\alpha  \left( {B_q^L P_L  + B_q^R P_R } \right)\mu \:,
\\&& B_q^{L,R}  = B_q^{(n)L,R}  + B_q^{(c)L,R}  + B_q^{(W)L,R}  (q=u,d) \:.
 \end{eqnarray}
$B_q(n)$ represent the contributions from the virtual neutral Fermion diagrams in the first line of Fig. \ref{FMFLV2}.
\begin{eqnarray}
 && B_q^L (n) = \sum_{i,j=1}^4\sum_{k,l=1}^6\Bigg\{\frac{1}{8e^2 m_W^2 }G_6 \left( x_{\chi _i^0 } ,x_{\chi _j^0 } ,x_{\tilde L_k} ,x_{\tilde Q_l}  \right)
\left[ {\left( H_R^{\tilde L_k\chi _i^0 \bar \mu } \right)}^* H_R^{\tilde L_k\chi _j^0 \bar e} H_R^{\tilde Q_l\chi _i^0 \bar q}
 {\left( H_R^{\tilde Q_l\chi _j^0 \bar q}  \right)}^*\right.
 \nonumber \\
 &&\hspace{2.8cm}\left. - {\left( H_R^{\tilde L_k\chi _i^0 \bar \mu }  \right)}^* H_R^{\tilde L_k\chi _j^0 \bar e}
  {\left( H_L^{\tilde Q_l\chi _i^0 \bar q}  \right)}^*H_L^{\tilde Q_l\chi _j^0 \bar q} \right]\nonumber \\
 &&\hspace{1.8cm} - \frac{{m_{\chi _i^0 } m_{\chi _j^0 } }}{{4e^2 m_W^2 }}G_7
 \left( x_{\chi _i^0 } ,x_{\chi _j^0 } ,x_{\tilde L_k} ,x_{\tilde Q_l}  \right)
 \left[ {\left( {H_R^{\tilde L_k\chi _i^0 \bar \mu } } \right)^* H_R^{\tilde L_k\chi _j^0 \bar e} H_L^{\tilde Q_l\chi _j^0 \bar q}
  \left( {H_L^{\tilde Q_l\chi _j^0 \bar q} } \right)^* } \right. \nonumber \\
 &&\hspace{2.8cm}\left. { - \left( {H_R^{\tilde L_k\chi _i^0 \bar \mu } } \right)^* H_R^{\tilde L_k\chi _j^0 \bar e}
  \left( {H_R^{\tilde Q_l\chi _j^0 \bar q} } \right)^* H_R^{\tilde Q_l\chi _j^0 \bar q} } \right]\Bigg\}, \nonumber \\
 &&B_q^R (n) = B_q^L (n)|_{L \leftrightarrow R} , \;\;\left( q = u,d \right).
 \end{eqnarray}

The virtual charged Fermion in the middle line of Fig. \ref{FMFLV2} give contributions denoted by $B_q(c)$.
\begin{eqnarray}
&& B_d^L (c) = \sum_{i,j=1}^2\sum_{k,l=1}^6\Bigg\{ \frac{1}{{8e^2 m_W^2 }}G_6 \left( {x_{\chi _i^ \pm  } ,x_{\chi _j^ \pm  } ,x_{\tilde \nu_k } ,x_{U_l^ +  } } \right)\left[ {\left( {H_R^{\tilde \nu_k \chi _i^ \pm  \bar \mu } } \right)^* H_R^{\tilde \nu_k \chi _j^ \pm  \bar e} H_R^{U_l^ +  \chi _i^ \pm  \bar d}
 \left( {H_R^{U_l^ +  \chi _j^ \pm  \bar d} } \right)^* } \right.  \nonumber \\
  &&\hspace{1.8cm}- \frac{{m_{\chi _i^ \pm  } m_{\chi _j^ \pm  } }}{{4e^2 m_W^2 }}G_7 \left( {x_{\chi _i^ \pm  } ,x_{\chi _j^ \pm  } ,x_{\tilde \nu_k } ,x_{U_l^ +  } } \right)
  \left. {\left( {H_R^{\tilde \nu_k \chi _i^ \pm  \bar \mu } } \right)^* H_R^{\tilde \nu_k \chi _j^ \pm  \bar e}
  H_L^{U_l^ +  \chi _i^ \pm  \bar d} \left( {H_L^{U_l^ +  \chi _j^ \pm  \bar d} } \right)^* } \right]  \Bigg\},\nonumber \\
&& B_u^L (c) =   \sum_{i,j=1}^2\sum_{k,l=1}^6\Bigg\{ \frac{-1}{{8e^2 m_W^2 }}
G_6 \left( {x_{\chi _i^ \pm  } ,x_{\chi _j^ \pm  } ,x_{\tilde \nu_k } ,x_{D_l^ -  } } \right)
\left[ {\left( {H_R^{\tilde \nu_k \chi _i^ \pm  \bar \mu } } \right)^* H_R^{\tilde \nu_k \chi _j^ \pm  \bar e}
 \left( {H_L^{D_l^ -  u^c \overline {\chi _i^ \pm  } } } \right)^* H_L^{D_l^ -  u^c \overline {\chi _j^ \pm  } } } \right. \nonumber\\
 &&\hspace{1.8cm} + \frac{{m_{\chi _i^ \pm  } m_{\chi _j^ \pm  } }}{{4e^2 m_W^2 }}G_7 \left( {x_{\chi _i^ \pm  } ,x_{\chi _j^ \pm  } ,x_{\tilde \nu_k } ,x_{D_l^ -  } } \right)
 \left. {\left( {H_R^{\tilde \nu_k \chi _i^ \pm  \bar \mu } } \right)^* H_R^{\tilde \nu_k \chi _j^ \pm  \bar e}
 \left( {H_R^{D_l^ -  u^c \overline {\chi _i^ \pm  } } } \right)^* H_R^{D_l^ -  u^c \overline {\chi _j^ \pm  } } } \right]\Bigg\}, \nonumber\\
 &&B_q^R (c) = B_q^L (c)|_{L \leftrightarrow R} \; .
\end{eqnarray}

The virtual $W$ produces corrections through the diagrams  in the last line of Fig. \ref{FMFLV2}
\begin{eqnarray}
 &&B_d^L (W) =  - \frac{1}{{2e^2 }}\frac{\partial }{{\partial x_W }}G_2 \left( {x_W ,x_\nu  ,x_u } \right)
 \left( {H_L^{W\mu \bar \nu } H_L^{W^* \nu \bar e} H_L^{W^* u\bar d} H_L^{Wd\bar u} } \right), \nonumber\\
 &&B_u^L (W) = \frac{2}{{e^2 }}\frac{\partial }{{\partial x_W }}G_2 \left( {x_W ,x_\nu  ,x_d } \right)
 \left( {H_L^{W^* \mu \bar \nu } H_L^{W\nu \bar e} H_L^{W^* d\bar u} H_L^{Wu\bar d} } \right), \nonumber\\
&&B_d^R (W) = B_u^R (W) = 0.
 \end{eqnarray}

\subsection{$\mu-e$ conversion rate }

Once we know the effective Lagrangian relevant to this process at the quark level, we can
calculate the conversion rate
\begin{eqnarray}
&&CR=4\alpha^5\frac{Z_{eff}^4}{Z}\left|F\left(q^2\right)\right|^2m_\mu^5\left[\left|Z\left(A_1^L-A_2^R\right)-\left(2Z+N\right)\bar{D}_u^L-\left(Z+2N\right)\bar{D}_d^L\right|^2\right.
\nonumber\\
&&\hspace{1.0cm}  \left.+\left|Z\left(A_1^R-A_2^L\right)-\left(2Z+N\right)\bar{D}_u^R-\left(Z+2N\right)\bar{D}_d^R\right|^2\right]\frac{1}{\Gamma_{capt}},
\\
&&\bar{D}_q^L=D_q^L+\frac{Z_L^q+Z_R^q}{2}\frac{F_L}{m_Z^2s_W^2c_W^2},
\nonumber\\
&&\bar{D}_q^R=\left.\bar{D}_q^L\right|_{{L}\leftrightarrow{R}}, ~~~(q=u,d).
\end{eqnarray}
with $Z$ and $N$ representing the proton and neutron numbers in a nucleus. $Z_{eff}$ is an
effective atomic charge determined in refs \cite{Zeff1,Zeff2}. $F\left(q^2\right)$ is the nuclear
form factor and $\Gamma_{capt}$ denotes the total muon capture rate, while $\alpha$ is the fine structure constant.
\section{numerical results}
In this section, we discuss the numerical results, and consider the experimental constraints from
the lightest neutral CP-even Higgs mass
$m_{_{h^0}}\simeq125.1\;{\rm GeV}$ \cite{LHC1,LHC2,LHC3} and the neutrino experiment data. In this model,
the LFV processes $l_j\rightarrow l_i\gamma$ and $l_j\rightarrow 3l_i$ are studied in our previous work\cite{weBLMSSM3},
 and their constraints are also taken into account.
In this work, we use the parameters\cite{weBLMSSM1,weBLMSSM2}
\begin{eqnarray}
&&L_4={3\over2},~~
 \mu=0.5 {\rm GeV}, ~~(A_l)_{ii}=-2000{\rm GeV},~~(A'_l)_{ii}=A'_L=300{\rm GeV}, \nonumber\\
&&(A_{N^c})_{ii}=(A_N)_{ii}=-500{\rm GeV},~~ (A_u)_{ii}=(A_d)_{ii}=(A'_d)_{ii}=(m_{\tilde{N}^c})_{ii}=1000{\rm GeV},\nonumber\\
&& (A'_u)_{ii}=800{\rm GeV},~~ (m_{\tilde{Q}})_{ii}=(m_{\tilde{U}})_{ii}=(m_{\tilde{D}})_{ii}=2\times10^3 {\rm GeV},~~\lambda_{N^c}=1.
\end{eqnarray}

The Yukawa couplings of neutrinos $(Y_\nu)^{IJ},(I,J=1,2,3)$ are at the order of $10^{-8}\sim10^{-6}$, whose effects to this processes are tiny and can be ignored savely.
 To simplify the numerical discussion, we use the following relations
  \begin{eqnarray}
 &&(A_l)_{ii}=AL,~~~~(A_{N^c})_{ii}=(A_{N})_{ii}=AN,~~~(A'_l)_{ii}=A'_L,\nonumber\\&&
(m^2_{\tilde{N}^c})_{ii}=M_{sn}^2,
~~~(m_{\tilde{L}}^2)_{ii}=(m_{\tilde{R}}^2)_{ii}=s_m^2, ~~~\texttt{for}~~ i=1,2,3.
\nonumber\\&&(m_{\tilde{L}}^2)_{ij}=(m_{\tilde{R}}^2)_{ij}=M_{Lf}, ~~~\texttt{for}~~ i,j=1,2,3 ~\texttt{and} ~i\neq j.
\end{eqnarray}
If we do not specially declare, the non-diagonal elements of the used parameters should be zero.

\subsection{$\mu\rightarrow e $ conversion rate in nuclei Au}
The experimental upper bound for the $\mu\rightarrow e $ conversion rate in nuclei Au is around
$7.0\times10^{-13}$. The parameters $ m_2=1000{\rm GeV}, M_{Lf}=10^4\, {\rm GeV^2}$
are supposed in the calculation of this process.
The parameter $m_1$ is related
to the mass matrix of the neutralino, which means
the contributions from neutralino-slepton can be influenced
by the parameter $m_1$.
For $S_m^2=13\,{\rm TeV^2}$ , ${\rm tan}\beta=5.0$, $\tan\beta_L=2.0$, and $g_L=\frac{1}{6}$, we plot the results versus $m_1$ with $V_{Lt}=3000\,{\rm GeV}\;{\rm and}\; 6000\,{\rm GeV}$ in Fig.3.
We can see that the results decrease quickly with the increase of $m_1$. As $V_{Lt}=6000\;{\rm GeV}$, the results are slightly smaller than the corresponding results with $V_{Lt}= 3000\,{\rm GeV}$.
This implies that $m_1$ is a sensitive parameter and has a strong effect on muon conversion to electron in nuclei.
Compared with $m_1$, the effect from $V_{Lt}$ is very small.

\begin{figure}[h]
\centering
\includegraphics[width=8cm]{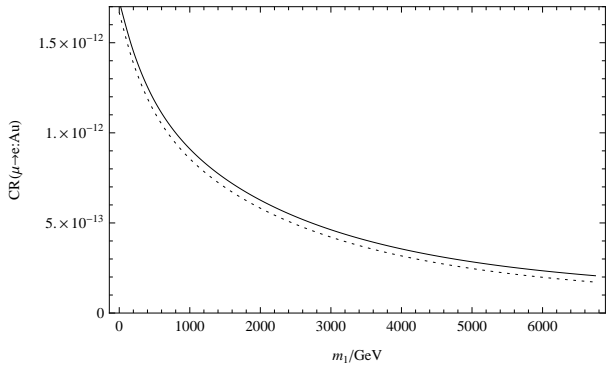}\\
\caption[]{With $ m_2=1000{\rm GeV}, M_{Lf}=10^4\, {\rm GeV^2}$,  $S_m^2=13\,{\rm TeV^2}$ , ${\rm tan}\beta=5.0$, $\tan\beta_L=2.0$, and $g_L=\frac{1}{6}$, $\mu\rightarrow e$ conversion rate in nuclei Au versus $m_1$ with $V_{Lt}=3000\,{\rm GeV}$ (solid line) and $ 6000\,{\rm GeV}$ (dotted line).}\label{fig3}
\end{figure}

${\rm tan}\beta$ is related to $v_u$ and $v_d$, and appears
in almost all mass matrices of particles contributing to the $\mu\rightarrow e $ processes.
With $ m_1=500 \,{\rm GeV}$, $V_{Lt}=3000\,{\rm GeV}$, $\tan\beta_L=2.0$, and $g_L=\frac{1}{6}$, Fig.4 shows
the variation of the $\mu\rightarrow e $ conversion rate in nuclei Au with the parameter
${\rm tan}\beta$ and $S_m^2$. It indicates that the results  change significantly with ${\rm tan}\beta$.
When ${\rm tan}\beta$ is in the region $(0\sim6)$, the results decrease significantly, but in the range of ${\rm tan}\beta>6$,
 we find that the results increase sharply. Only when the value of ${\rm tan}\beta$  is about 6, the results of  $\mu\rightarrow e $ conversion rate in nuclei Au are close and not higher than the experimental upper bound.
\begin{figure}[h]
\centering
\includegraphics[width=8cm]{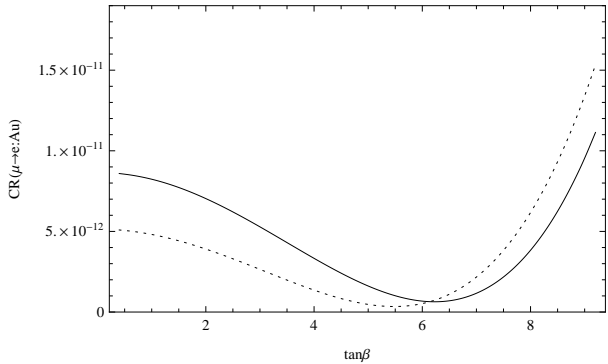}\\
\caption[]{With $ m_1=500 \,{\rm GeV}$, $ m_2=1000\,{\rm GeV}, M_{Lf}=10^4\, {\rm GeV^2}$, $V_{Lt}=3000\,{\rm GeV}$, $\tan\beta_L=2.0$, and $g_L=\frac{1}{6}$, the contribution to  $\mu\rightarrow e $ conversion rate in nuclei Au versus ${\rm tan}\beta$ with $S_m^2=12 \,{\rm TeV^2}$ (solid line) and   $16 \,{\rm TeV^2}$(dotted line).} \label{fig4}
\end{figure}

The parameters  $g_L$, $\tan\beta_L$ and $V_{Lt}$ all present in the mass squared matrices of sleptons,
sneutrinos and lepton neutralinos. Therefore, these three parameters affect the results
through slepton-neutrino, sneutrinos-chargino and slepton-lepton neutralino contributions.

As $ m_1=1000{\rm GeV},\tan\beta=5.5, \tan\beta_L=2,S_m^2=16\,{\rm TeV^2}$,  $g_L$ versus $V_{Lt}$ are scanned in Fig.5. We
find that the allowed scope of $V_{Lt}$ shrinks and the value of  $V_{Lt}$ decreases with the enlarging
$g_L$. Therefore, the value of $g_L$ should not be too large. Generally, we take $0.05 \leq g_L \leq 0.3$ and
$V_{Lt}\sim 3\,$TeV in our numerical calculations.

\begin{figure}[h]
\centering
\includegraphics[width=8cm]{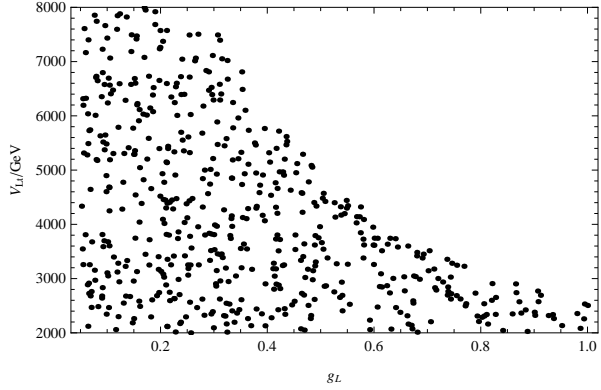}\\
\caption[]{For $\mu\rightarrow e $ conversion rate in nuclei Au, the allowed parameter space in the plane of  $g_L$ versus $V_{Lt}$ with $ m_1=1000 \,{\rm GeV} $, $ m_2=1000\,{\rm GeV}$, ${\rm tan}\beta=5.5$, $S_m^2=16\,{\rm TeV^2}$, $ M_{Lf}=10^4\, {\rm GeV^2}$ and $\tan\beta_L=2$.  \label{fig5}}
\end{figure}

As the parameters $m_1=1000\,{\rm GeV}$, ${\rm tan}\beta=6.0$, $S_m^2=12\,{\rm TeV^2}$ and $V_{Lt}=3000\,{\rm GeV}$, we plot the allowed results with $\tan\beta_L$ versus $g_L$  in Fig.6. When $g_L<0.43$, the parameter ${\rm tan}\beta_L$ can vary in the region of $(0\sim 4)$. It implies that  $g_L$ is a sensitive parameter to the numerical results and the value of $g_L$ should not be larger than 0.43.

\begin{figure}[h]
\centering
\includegraphics[width=8cm]{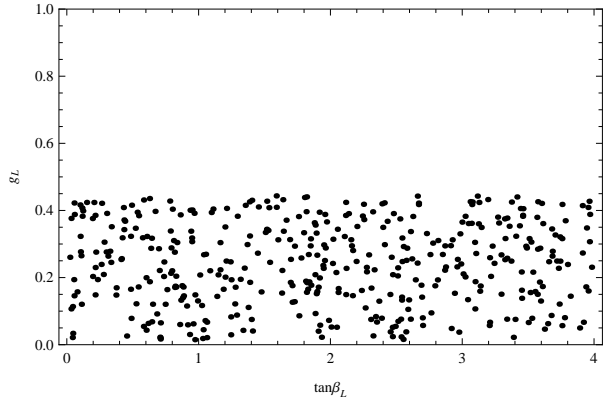}\\
\caption[]{For $\mu\rightarrow e $ conversion rate in nuclei Au, the allowed parameter space in the plane of $\tan\beta_L$ versus  $g_L$ with $ m_1=1000 \,{\rm GeV} $, $ m_2=1000\,{\rm GeV}$, ${\rm tan}\beta=6 $, $S_m^2=12{\rm TeV^2}$,  $ M_{Lf}=10^4\, {\rm GeV^2}$ and $V_{Lt}=3000\,{\rm GeV}$.   \label{fig6}}
\end{figure}

\subsection{$\mu\rightarrow e $ conversion rate in nuclei Ti}
In a similar way, the $\mu\rightarrow e $ conversion rate in nuclei Ti is numerically
studied and its experimental upper bound is around
$4.3\times10^{-12}$. In this subsection, we use the parameters ${\rm tan}\beta=2.0$,  $\tan\beta_L=2.0$, $g_L=\frac{1}{6}$ and $V_{Lt}=3000\,{\rm GeV}$.
The parameter $m_2$ presents
in the mass matrixes of neutralino and chargino. This parameter affects the numerical
results through the neutralino-slepton and chargino-sneutrino
contributions.
With $ S_m^2=10\,{\rm TeV^2}, M_{Lf}=10^4\, {\rm GeV^2}$, we plot the results versus $m_2$ with  $ m_1=1000\,{\rm GeV}$ and $ 2000\,{\rm GeV}$ by the solid and dotted lines in Fig.7. We can see that the results decrease with the increase of $m_2$. The results of dotted line are slightly larger than solid line and all the results are in the region $(1.4 \times10^{-12}\sim6.4 \times10^{-12})$. This implies that $m_2$ should have impact on the results to some extent.

\begin{figure}[h]
\centering
\includegraphics[width=8cm]{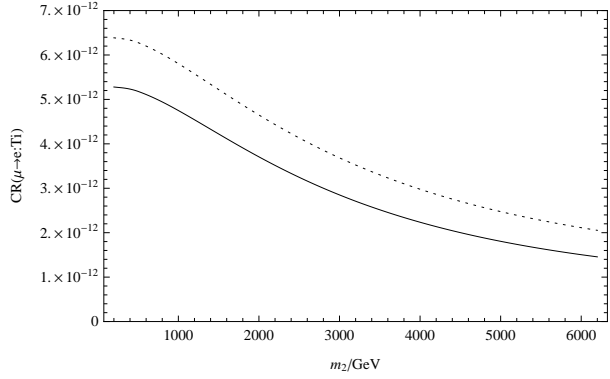}\\
\caption[]{ With ${\rm tan}\beta=2.0$, $S_m^2=10\,{\rm TeV^2}, M_{Lf}=10^4\, {\rm GeV^2}$, $\tan\beta_L=2.0$, $g_L=\frac{1}{6}$ and $V_{Lt}=3000\,{\rm GeV}$, $\mu\rightarrow e $ conversion rate in nuclei Ti versus $m_2$ is plotted for $ m_1=2000\,{\rm GeV}$ (solid line) and $1000\,{\rm GeV}$ (dotted line).} \label{fig7}
\end{figure}
$M_{Lf}$ are the non-diagonal elements of ${m_{\tilde{L}}}^2$ and ${m_{\tilde{R}}}^2$ in the slepton mass matrix.
 For $m_1=1000\,{\rm GeV}$ and $S_m^2=12\,{\rm TeV^2}$, we study the $\mu\rightarrow e $ conversion rate in nuclei Ti
versus $M_{Lf}$ with $m_2=1000\, {\rm GeV}$ (solid line) and  $m_2=2000\, {\rm GeV}$  (dotted line) in Fig.8.
As $M_{Lf}=0$, the conversion ratio for  $\mu\rightarrow e $ is almost
zero, but the results increase sharply with $M_{Lf}>0$. We deduce that non-zero $M_{Lf}$ is a sensitive parameter and has a strong effect on muon conversion to electron in nuclei.

\begin{figure}[h]
\centering
\includegraphics[width=8cm]{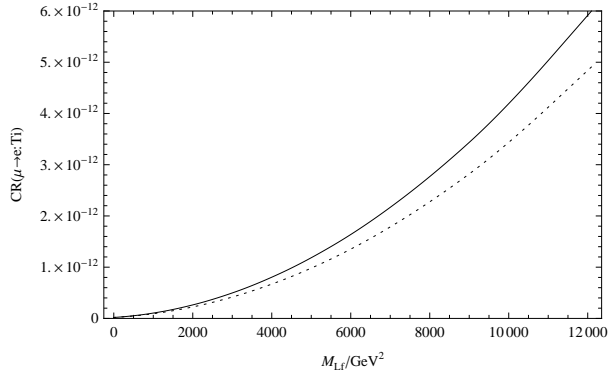}\\
\caption[]{With $ m_1=1000\, {\rm GeV}$, ${\rm tan}\beta=2.0$, $S_m^2=12\,{\rm TeV^2}$, $\tan\beta_L=2.0$, $g_L=\frac{1}{6}$ and $V_{Lt}=3000\,{\rm GeV}$, $\mu\rightarrow e $ conversion rate in nuclei Ti versus $M_{Lf}$ with $m_2=1000 \,{\rm GeV}$ (solid line) and 2000 \,{\rm GeV}(dotted line).} \label{fig8}
\end{figure}

\subsection{$\mu\rightarrow e $ conversion rate in nuclei Pb}
The experimental upper bound of $\mu\rightarrow e $ conversion rate in nuclei Pb is around
$4.6\times10^{-11}$. In this subsection, we use the parameters $m_2=1000\,{\rm GeV}, \tan\beta_L=2.0$, $g_L=\frac{1}{6}$ and $M_{Lf} =10^4\, {\rm GeV^2}$.
$S_m$ are the diagonal elements of ${m_{\tilde{L}}}^2$ and ${m_{\tilde{R}}}^2$
in the slepton mass matrix,
which can affect slepton-neutralino and slepton-lepton
neutralino contributions in the $\mu\rightarrow e $ process.

With $m_1=1000 \,{\rm GeV}, V_{Lt}=3000\, {\rm GeV}$, we plot the conversion ratio for  $\mu\rightarrow e $ in nuclei Pb versus
$S_m$ with ${\rm tan} \beta=2.0$ (solid line) and  ${\rm tan} \beta=3.0$ (dotted line)  in Fig.9. These two
lines decrease quickly with $S_m$ enlarging from 1400 GeV to 3000 GeV, which indicates that
$S_m$ is a very sensitive parameter to the numerical results. When $S_m > 3000\, {\rm GeV}$, the results
decrease slowly and the conversion ratios are around ($10^{-12}\sim10^{-13}$).

\begin{figure}[h]
\centering
\includegraphics[width=8cm]{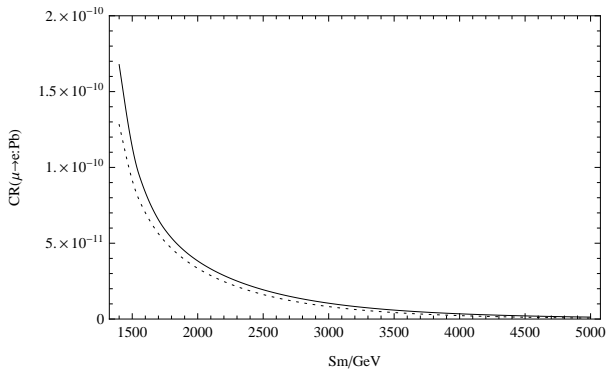}\\
\caption[]{With  $m_1=1000 \,{\rm GeV}$, $m_2=1000 \,{\rm GeV}$, $M_{Lf} =10^4 \,{\rm GeV^2}$, ${\rm tan}{\beta_L}=2.0$, $g_L=\frac{1}{6} $ and $V_{Lt}=3000\, {\rm GeV}$, the contribution to  $\mu\rightarrow e $ conversion rate in nuclei Pb versus $S_m$ with ${\rm tan} \beta=2 $ (solid line) and 3 (dotted line).} \label{fig9}
\end{figure}

We focus on $V_{L_t}$ which is a special parameter in BLMSSM, and with $m_1=500 \,{\rm GeV},\, {\rm tan} \beta=13$, we plot the conversion ratio for  $\mu\rightarrow e $ in nuclei Pb versus $V_{L_t}$ with $S_m^2=5\,{\rm TeV^2}$ (solid line) and $S_m^2=6\,{\rm TeV^2}$ (dotted line)  in Fig. 10. Overall, the results of dotted line are about $0.5\times10^{-11}\sim1.2\times10^{-11}$ larger than the solid line. In the range of $V_{L_t}=(0\sim10000\,{\rm GeV})$, the two lines decrease quickly with the enlarging $V_{L_t}$. We can see $S_m$ and $V_{L_t}$ are sensitive parameters to the numerical results.

\begin{figure}[h]
\centering
\includegraphics[width=8cm]{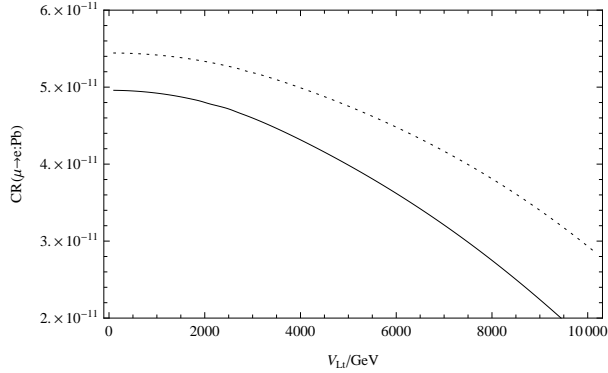}\\
\caption[]{With $ m_1=500 \,{\rm GeV},\, m_2=1000 \,{\rm GeV},\, {\rm tan} \beta=13,\, M_{Lf} =10^4 \,{\rm GeV^2}$, ${\rm tan} \beta_L=2.0 $, and $g_L=\frac{1}{6}$, $\mu\rightarrow e $ conversion rate in nuclei Pb versus $V_{Lt}$ with $S_m^2=5\,{\rm TeV^2}$ (solid line) and $S_m^2=6\,{\rm TeV^2}$ (dotted line).} \label{fig10}
\end{figure}

\section{discussion and conclusion}
In the framework of the BLMSSM model, we study the LFV processes $\mu\rightarrow \emph{e}+\rm{q}\bar q$. In the processes,
we consider some new parameters and contributions, such as the newly introduced parameters $g_L$, $\tan{\beta_L}$ and $V_{L_t}$.
Combined with the numerical results discussed in the Section IV, different parameters have different effects on the processes.
The parameter  $g_L$ presents in the mass squared matrices of sleptons,
sneutrinos and lepton neutralinos. Numerical analysis shows that  $g_L$ has obvious influence on the results, the value of $g_L$ should not be too large.
  As sensitive parameters, $S_m$ and $M_{L_f}$ are respectively diagonal and non-diagonal elements of matrixes for $m_{\tilde{L}}$ and $m_{\tilde{R}}$.
   Both $S_m$ and $M_{L_f}$  have  significant impacts on the results. ${\rm tan}\beta$ is related to $v_u$ and $v_d$, and appears
in almost all mass matrices of particles contributing to the $\mu\rightarrow e $ processes. The value of ${\rm tan}\beta$
is critical to these processes.
 With the improvement of experimental accuracy, we believe that there will be some discoveries for $\mu$ to e conversion in the near future.
\section{Acknowledgments}
Supported by National Natural Science Foundation of China (No. 11535002 and No.11575052),
Physics laboratory center of Hebei GEO University(Hebei experimental teaching demonstration center).
\section{Appendix }
In this section, we give out the corresponding one loop functions.
${G_2}(\textit{x}_1 , x_2 , x_3)$ and ${G_3}(\textit{x}_1 , x_2 , x_3)$ have infinite term, and to obtain finite results we use $\overline{MS}$ subtraction and $\overline{DR}$ scheme.

\begin{eqnarray}
&&{I_1}(\textit{x}_1 , x_2 ) = \frac{1}{{96{\pi ^2}}} \Big[ \frac{{11 + 6\ln {x_2}}}{{({x_2} - {x_1})}}- \frac{{15{x_2} + 18{x_2}\ln {x_2}}}{{{{({x_2} - {x_1})}^2}}} + \frac{{6x_2^2 + 18x_2^2\ln {x_2}}}{{{{({x_2} - {x_1})}^3}}},  \nonumber\\
&&\qquad\qquad\quad\; + \: \frac{{6x_1^3\ln {x_1}}-{6x_2^3\ln {x_2}}}{{{{({x_2} - {x_1})}^4}}}  \Big]\:.\\
&&{I_2}(\textit{x}_1 , x_2 ) = \frac{1}{{32{\pi ^2}}}\Big[  \frac{{3 + 2\ln {x_2}}}{{({x_2} - {x_1})}} - \frac{{2{x_2} + 4{x_2}\ln {x_2}}}{{{{({x_2} - {x_1})}^2}}} -\frac{{2x_1^2\ln {x_1}}}{{{{({x_2} - {x_1})}^3}}}+ \frac{{2x_2^2\ln {x_2}}}{{{{({x_2} - {x_1})}^3}}}\Big]\:, \\&&{I_3}(\textit{x}_1 , x_2 ) = \frac{1}{{16{\pi ^2}}}\Big[ \frac{{1 + \ln {x_2}}}{{({x_2} - {x_1})}} + \frac{{{x_1}\ln {x_1}}-{{x_2}\ln {x_2}}}{{{{({x_2} - {x_1})}^2}}} \Big]\:.
\\&&
{I_4}(\textit{x}_1 , x_2 ) = \frac{1}{{16{\pi ^2}}}\Big[ - \frac{{1 + \ln {x_1}}}{{({x_2} - {x_1})}} - \frac{{{x_1}\ln {x_1}}-{{x_2}\ln {x_2}}}{{{{({x_2} - {x_1})}^2}}} \Big]\:.
\\
&&{G_1}(\textit{x}_1 , x_2 , x_3) =  \frac{1}{{16{\pi ^2}}}\Big[ \frac{{{x_1}\ln {x_1}}}{{({x_1} - {x_2})({x_1} - {x_3})}} + \frac{{{x_2}\ln {x_2}}}{{({x_2} - {x_1})({x_2} - {x_3})}} \nonumber\\
&&\qquad\qquad\qquad\quad +  \: \frac{{{x_3}\ln {x_3}}}{{({x_3} - {x_1})({x_3} - {x_2})}}\Big], \\
&&{G_2}(\textit{x}_1 , x_2 , x_3) =  \frac{1}{{16{\pi ^2}}}\Big[ -(\Delta  + 1 + \ln {x_\mu })  + \frac{{x_1^2\ln {x_1}}}{{({x_1} - {x_2})({x_1} - {x_3})}}  \nonumber\\
&&\qquad\qquad\qquad\quad  + \: \frac{{x_2^2\ln {x_2}}}{{({x_2} - {x_1})({x_2} - {x_3})}}+ \frac{{x_3^2\ln {x_3}}}{{({x_3} - {x_1})({x_3} - {x_2})}} \Big].
\end{eqnarray}

\begin{eqnarray}
&&G_3(x_1,x_2)=\frac{-1}{16 \pi ^2}\Big((\Delta+\ln
   x_\mu+1)+\frac{x_2^2
   \ln x_2-x_1^2 \ln x_1}{(x_2-x_1)^2}+\frac{x_2+2 x_2 \ln
   x_2}{x_1-x_2}-\frac{1}{2}\Big),\nonumber\\&&
G_4(x_1,x_2,x_3)=\frac{1}{32\pi^2}\Big(\frac{2x_1^3[3x_1(x_1-x_2-x_3)+x_2^2+x_2x_3+x_3^2]\ln x_1}{(x_1-x_2)^3(x_1-x_3)^3}\nonumber\\&&\hspace{2.6cm}-\frac{2(3x_1^2-3x_1x_2+x_2^2)x_2\ln x_2}{(x_1-x_2)^3(x_2-x_3)}+\frac{2(3x_1^2-3x_1x_3+x_3^2)x_3\ln x_3}{(x_1-x_3)^3(x_2-x_3)}\nonumber\\&&\hspace{2.6cm}
-\frac{x_1[5x_1^2-3x_1(x_2+x_3)+x_2x_3]}{(x_1-x_2)^2(x_1-x_3)^2}
\Big),\nonumber\\
&&G_5(x_1,x_2,x_3)=\frac{1}{16\pi^2}\Big(\frac{x_1^2(2x_1-x_2-x_3)\ln x_1}{(x_1-x_2)^2(x_1-x_3)^2}+\frac{x_2(x_2-2x_1)\ln x_2}{(x_1-x_2)^2(x_2-x_3)}
\nonumber\\&&\hspace{2.6cm}-\frac{x_1}{(x_1-x_2)(x_1-x_3)}+\frac{x_3(2x_1-x_3)\ln x_3}{(x_1-x_3)^2(x_2-x_3)}\Big).
\end{eqnarray}
\begin{eqnarray}
&&{G_6}(\textit{x}_1 , x_2 , x_3, x_4) = \frac{1}{{16{\pi ^2}}}\Big[\frac{{x_1^2\ln {x_1}}}{{({x_1} - {x_2})({x_1} - {x_3})({x_1} - {x_4})}}  +\; \frac{{x_2^2\ln {x_2}}}{{({x_2} - {x_1})({x_2} - {x_3})({x_2} - {x_4})}}\nonumber\\
&&\hspace{3.0cm}  + \frac{{x_3^2\ln {x_3}}}{{({x_3}  - {x_1})({x_3} - {x_2})({x_3} - {x_4})}} + \frac{{x_4^2\ln {x_4}}}{{({x_4} - {x_1})({x_4} - {x_2})({x_4} - {x_3})}}\Big]\:,
\nonumber\\&&
{G_7}(\textit{x}_1 , x_2 , x_3, x_4) = \frac{1}{{16{\pi ^2}}}\Big[\frac{{{x_1}\ln {x_1}}}{{({x_1} - {x_2})({x_1} - {x_3})({x_1} - {x_4})}}  +\frac{{{x_2}\ln {x_2}}}{{({x_2} - {x_1})({x_2} - {x_3})({x_2} - {x_4})}} \nonumber\\
&&\hspace{3.2cm} + \frac{{{x_3}\ln {x_3}}}{{({x_3}  - {x_1})({x_3} - {x_2})({x_3} - {x_4})}}  + \: \frac{{{x_4}\ln {x_4}}}{{({x_4} - {x_1})({x_4} - {x_2})({x_4} - {x_3})}}\Big] .
\end{eqnarray}

 \end{document}